\documentclass[epj]{svjour}
\usepackage{graphics}
\begin{document}
\title{An Inflationary Scenario Taking into Account of Possible \\Dark Energy Effects in the Early Universe}
\author{Zhe Chang\inst{1,2} \and Ming-Hua Li\inst{1,2} \and Sai Wang\inst{1,2} \and Xin Li\inst{1,2}
}                     
%
%
\institute{Institute of High Energy Physics, Chinese Academy of Sciences, 100049 Beijing, China \and Theoretical Physics Center for Science Facilities, Chinese Academy of Sciences, 100049 Beijing, China}
\date{Received: date / Revised version: date}
%
\abstract{
We investigate the possible effect of cosmological-constant type dark energy during the inflation period of the early universe. This is accommodated by a new dispersion relation in de Sitter space. The modified inflation model of a minimally-coupled scalar field is still able to yield an observation-compatible scale-invariant primordial spectrum, simultaneously having potential to generate a spectrum with lower power at large scales. A qualitative match to the WMAP 7-year data is presented. We obtain an $\Omega_\Lambda$ of the same order of that in the $\Lambda$-CDM model. Possible relations between the de Sitter scenario and the Doubly Special Relativity(DSR) are also discussed.
} 
\maketitle
\section{Introduction}
\label{intro}
The anisotropy of the Cosmic Microwave Background(CMB) radiation, which was first discovered by the NASA Cosmic Background Explorer(COBE) satellite in the early 1990s, has been confirmed by subsequent balloon experiments and more recently by the Wilkinson Microwave Anisotropy Probe(WMAP)'s 7-year results \cite{wmap7}\cite{0067-0049-192-2-18}. The temperature fluctuations of the CMB are believed by most cosmologists to be generated as quantum fluctuations of a weakly self-coupled scalar matter field $\phi$ , which later leads to the exponential inflation of the early universe \cite{Star}-\cite{Mukh2}. The WMAP 7-year results provide a strong confirmation of the inflationary paradigm. With the prestigious $\Lambda$-CDM model (also known as the ``concordance model'' or the ``cosmological standard model'') and the basic set of six cosmological parameters \cite{H2010-}\cite{Larson2011-}, one can now give a unprecedently good global description of the universe, with an accuracy down to $10\%$ level.

However, various ``anomalies'' have been reported about the CMB data. One of them is the low-$\ell$ multipole controversy \cite{J2003-small}\cite{Efst}\cite{Kurki-Suonio2010}. It was reported as ``smaller than the standard model-predicted'' observed values of $C_{\ell}$ for low $\ell$, especially for the quadrupole component $\ell =2$. This issue had been widely investigated in the last decade \cite{Kowaski2003-,Feng2003-,C2003-Suppressing,Y2005-Possible,Kofman:1989ed,Hodges:1989dw,Rodrigues2008-,Copi2010}. To one's surprise, H. Liu and T.-P. Li \cite{Li2010-}\cite{Li2011-} claimed that the CMB quadrupole released by the WMAP team is almost completely artificial and the real quadrupole of the CMB anisotropy should be near zero. C. Bennett \textit{et al.} \cite{Bennett2011} had recently examined the properties of the power spectrum data from the WMAP 7-year release with respect to the $\Lambda$-CDM model carefully. On the contrary, they reported that the amplitude of the quadrupole is well within the expected $95\%$ confidence range and therefore is not anomalously low.

Considering that the WMAP's results have shown strong substantiating evidence for the concordance cosmological model, the remarkable agreement between the theory and the observational data should not be taken lightly. But in consideration of the work listed above, it may be prudent for one to leave the low-$\ell$ multipole issue as an open question for further investigations and observations.

Besides, there is another problem needed to be considered in most of the inflation models. That is the ``trans-Planckian'' problem. Since the inflation period has to lasts for sufficiently long to generate a $60$ to $65$ e-foldings number in order to solve the flatness and horizon problem of the universe, the wavelengths corresponding to the large-scale structure at present must be once smaller than the Planck length $\ell_{P}$, for which these theories break down. Similar problem appears in black hole physics. The calculations of Hawking radiation would become irrational if one traces the modes infinitely to the past. To address this issue, two major approaches are proposed by the black hole physicists. One is to apply the stringy space-time uncertainty relation \cite{Ho} on the fluctuation modes to pose a modification of the boundary condition \cite{Brand2003-}. The other is to mimic the quantum gravity effects by replacing the linear dispersion relation $\omega^2 =k^2$ (for photons) by a non-standard one that derived from a quantum gravity theory \cite{J2001-Trans-Planckian}.

On the other hand, some cosmologists believe that our universe in its early history can actually be approximated by a de Sitter one. The connection between a de Sitter space and the early universe lies on one or more of the hottest issues in modern cosmology---the cosmological constant and dark energy \cite{Peebles2003-}. The cosmological constant $\Lambda$, also taken as one form of dark energy, gives rise to the famous $\Lambda$-CDM model. Although the universe after inflation is described well by the standard cosmological model, the physical implications of a $\Lambda$-type dark energy during the inflation period is rarely discussed. Its possible effects are worth considering for a unified scenario of cosmological theory. Moreover, a de Sitter universe is a cosmological solution to Einstein's field equations of general relativity with a positive cosmological constant $\Lambda$. It models our universe as a spatially flat one and neglects ordinary matter. The dynamics of the universe are dominated by the cosmological constant $\Lambda$, which is thought to correspond to dark energy in our universe. All these give a physically befitting description of the universe at about a time $t = 10^{-33}$ seconds after the fiducial Big Bang.

In this paper, we take into account the effect of a cosmological-constant type dark energy during the inflation period in the early universe. We try to construct a unified scenario of the $\Lambda$-CDM and the inflation model. This is accommodated by a new dispersion relation---a dispersion relation in de Sitter space. It stems from the kinematics of free particles in a four dimensional de Sitter space. The CMB TT spectrum under the influence of such a form of dark energy during inflation is presented. We find that for certain parameter values, the modified inflation model yields an angular spectrum with lower power at large scales. An $\Omega_\Lambda$ of the same order of that in the $\Lambda$-CDM model is obtained. The relation with the Doubly Special Relativity(DSR) is also discussed. All the numerical results in this paper are qualitative. A full Bayesian analysis of the data like that carried by the WMAP team is necessary to prove that the model is actually statistically more consistent with the observations \cite{Bennett2011}.

The rest of the paper is organized as follows. In Section \ref{sec:1}, a remarkable dispersion relation in a four dimensional de Sitter space is introduced to investigate possible effects of the $\Lambda$-type dark energy on the single-scalar-field inflation model. In Section \ref{sec:2}, we obtain the corresponding primordial spectrum in the modified inflationary scenario. In Section \ref{sec:3}, we compare the corresponding CMB angular power spectrum with the WMAP 7-year data. An observation-compatible result has been obtained in a qualitative manner. Conclusions and discussions are presented in Section \ref{sec:4}. The relation between our model and the Doubly Special Relativity(DSR) is discussed in this section.

\section{Dispersion Relation in de Sitter Space}
\label{sec:1}
De Sitter space, first discovered by Willem de Sitter in the 1920s, is a maximally symmetric space in mathematics. It is a space with constant positive curvature. In the language of general relativity, de Sitter space is the maximally symmetric, vacuum solution of Einstein's field equations with a positive (or physically repulsive) cosmological constant $\Lambda$. It corresponds to a positive vacuum energy density with negative pressure of our universe, i.e. one form of dark energy.


A four dimensional de Sitter space (three space dimensions plus one time dimension) describes a cosmological model for the physical universe.
It can be realized as a four-dimensional pseudo-sphere imbedded in a five dimensional Minkowski flat space with coordinates $\xi_{\mu}$ ($\mu =0,1,2,3,4$), to wit \cite{S2004-Cosmic}\cite{G2003-Beltrami}
\begin{equation}
\label{xi}
\begin{array}{l}
\displaystyle
-\xi_0^2 + \xi_1^2 + \xi_2^2 +\xi_3^2 +\xi_4^2=\frac{1}{K}=R^2 \ ,\\[5mm]
ds^2=-d\xi_0^2 +d\xi_1^2 +d\xi_2^2 +d\xi_3^2 +d\xi_4^2\ ,
\end{array}
\end{equation}
where $K$ and $R$ respectively denotes the Riemannian curvature and radius of the de Sitter spacetime.
For mathematical reasons, we adopt the \emph{Beltrami coordinates} given by
\begin{equation}\label{xi-x}
x_{\mu}\equiv R\frac{\xi_{\mu}}{\xi_4} \ ,~~~~~\mu =0,1,2,3,~\textmd{and}~\xi_4\not=0 \,.
\end{equation}
In the Beltrami de Sitter(BdS) space, the line element in (\ref{xi}) can now be rewritten as
\begin{equation}
\begin{array}{l}
\sigma\equiv\sigma(x,~x)=1-K\eta_{\mu \nu}x^\mu x^\nu (>0) \ ,\\[3mm]
\displaystyle ds^2=\left(\frac{\eta^{\mu \nu}}{\sigma}+\frac{K \eta^{\mu \alpha}\eta^{\nu \beta}x_\alpha x_\beta}
{\sigma^2}\right)dx_\mu dx_\nu \ ,~~~~~\mu,\nu=0,1,2,3\ ,
\end{array}
\end{equation}
where $\eta_{\mu \nu}={\rm diag}(-1,+1,+1,+1)$ is the Minkowski metric.

The five dimensional angular momentum $M_{\mu \nu}$ of a free particle with mass $m_0$ is defined as
\begin{equation}\label{angular}
M_{\mu \nu}\equiv m_0\left(\xi_{\mu}\frac{d\xi_{\nu}}{ds}-\xi_{\nu}\frac{d\xi_{\mu}}{ds}\right)\ ,~~~~~\mu =0,1,2,3,4\ ,
\end{equation}
where $s$ is the affine parameter along the geodesic. In the de Sitter spacetime, there is no translation invariance so that one can not introduce a momentum vector. However, it should be noticed that, at least somehow, we may define a counterpart of the four dimensional momentum $P_\mu$ of a free particle in the de Sitter spacetime:
\begin{equation}
\label{momentum}
P_\mu \equiv R^{-1}M_{4\mu}=m_0 \sigma^{-1}\frac{dx_{\mu}}{ds}\ ,~~~~~\mu,\nu=0,1,2,3\, .
\end{equation}
For the rest of the article, the Greek indices (i.e. $\mu$, $\nu$, $\alpha$, $\beta$, etc.), if not specifically pointed out, run from $0$ to $3$. The Latin indices  (i.e. $i$, $j$, $k$, etc.) run from $1$ to $3$.

In the same manner, the counterparts of the four dimensional angular momentum $J_{\mu \nu}$ can be assigned as
\begin{equation}
\label{angular-mn} J_{\mu \nu}\equiv M_{\mu \nu} =
x_\mu P_\nu-x_\nu P_\mu=m_0\sigma^{-1}\left(x_\mu \frac{dx_\nu}{ds}-x_\nu \frac{dx_\mu}{ds}\right)\, .
\end{equation}
Under the de Sitter transformations for a free particle, an invariant (which turns out to be just ${m^2_0}$) can be constructed in terms of the angular momentum $M^{\mu\nu}$ as
\begin{equation}
\begin{array}{c}
\label{casimir}
m^2_0=\displaystyle\frac{\lambda}{2}M_{\mu\nu}M^{\mu\nu}=E^2-{\bf P}^2+\frac{K}{2} J_{ij}J^{ij}~,\\ [0.5cm]
E=P_0~,~~~~{\bf P}=(P_1,~P_2,~P_3)~.
\end{array}
\end{equation}

To discuss the quantum kinematics of a free particle, it is natural to realize the five dimensional angular momentum $M_{\mu\nu}$ as the infinitesimal generators $\hat{M}_{\mu\nu}$ of the de Sitter group $SO(1,4)$, to wit
\begin{equation}
\label{L-operator}
\hat{M}_{\mu\nu}\equiv-i\left(\xi_{\mu}\frac{\partial}{\partial\xi^{\nu}}-\xi_{\nu}
\frac{\partial}{\partial\xi^{\mu}}\right)\ ,~~~~~\mu,\nu=0,1,2,3\, .
\end{equation}

Since $\xi_0(\equiv\sigma(x,x)^{-1/2}x_0)$ is invariant under the spatial transformations of $x_\alpha$ acted by the subgroup $SO(4)$ of $SO(4,1)$, two space-like events are considered to be \emph{simultaneous} if they satisfy
\begin{equation}
\label{simultaneous} \sigma(x,~x)^{-\frac{1}{2}}x^0=\xi^0={\rm
constant}~.
\end{equation}
Therefore, it is convenient to discuss physics of the de Sitter spacetime in the coordinate $(\xi^0,~x^\alpha)$.
With the de Sitter invariant (or the Casimir operator) in the relation (\ref{casimir}), one can write down the \emph{equation of motion of a free scalar particle  with mass $m_0$ in the de Sitter space} as
\begin{equation}
\label{motion}
\left(\frac{K}{2}\hat{M}_{\mu\nu}\hat{M}^{\mu\nu}-m^2_0\right)\phi(\xi_0,x_{\alpha})=0~,
\end{equation}
where the $\phi(\xi_0,x_{\alpha})$ denotes the scalar field.

A laborious but straight forward process of solving the equation (\ref{motion}) can be found in \cite{S2004-Cosmic}, from which one obtains the \emph{dispersion relation} for a free scalar particle in the de Sitter space:
\begin{equation}
\label{dispersion}
E^2=m_0^2+\varepsilon'^2+K(2n+l)(2n+l+2)\ ,
\end{equation}
where $n$ and $l$ refer to the radial and the angular quantum number respectively. For massless particles such as photons, the above dispersion relation becomes
\begin{equation}
\begin{array}{c}
\label{dispersion1}
\omega^2= k^2+\varepsilon^{*2}_{\gamma}\ ,\\ [0.5cm]
\varepsilon^{*}_{\gamma}\equiv \sqrt{K(2n_\gamma +l_\gamma)(2n_\gamma +l_\gamma +2)}\ ,
\end{array}
\end{equation}
where $n_{\gamma}$ and $l_{\gamma}$  respectively denotes the radial and the angular quantum number. $w$ and $k$ are in turn the frequency and the wavenumber.

\section{The Modified Primordial Spectrum}
\label{sec:2}
Given the dispersion relation (\ref{dispersion1}), we are now in the position to calculate the primordial power spectrum ${\cal P}_{\delta \phi}(k)$, from which later the anisotropy spectrum of the CMB is obtained.

Let us denote the inhomogeneous perturbation to the inflaton field by $\delta\phi({\bf x},t)$. In the Fourier representation, the evolution equation of the primordial perturbation $\delta\sigma_{\bf k}$, which is defined as $\delta\sigma_{\bf k} \equiv a\delta\phi_{\mathbf{k}}$, reads \cite{A2002-}
\begin{equation}
\delta\sigma^{\prime\prime}_{\bf k} +\left(k^2_{\textmd{eff}}-\frac{2}{\tau^2}\right)\delta\sigma_{\bf k}=0\ ,
~~~~~k^2_{\textmd{eff}}\equiv k^2+\varepsilon^{*2}_{\gamma}~.
\label{KGeq5}
\end{equation}
A prime indicates differentiation with respect to the conformal time $\tau$ \cite{Dodelson2003-}.

The equation (\ref{KGeq5}) has an exact particular solution:
\begin{eqnarray}
\label{sigma}
\delta\sigma_{\bf k} &=& \frac{e^{-ik_{\textmd{eff}}\tau}}{\sqrt{2k_{\textmd{eff}}}}\left(
1+\frac{i}{k_{\textmd{eff}}\tau}\right) \nonumber \\
&=& \frac{e^{-i\sqrt{k^2+\varepsilon^{*2}_{\gamma}}~\tau}}{\sqrt{2}\left(k^2+\varepsilon^{*2}_{\gamma}\right)^{1/4}}
\left(1+ \frac{i}{\sqrt{k^2+\varepsilon^{*2}_{\gamma}}~\tau}\right)\ .
\end{eqnarray}
For the redefinition $\delta \phi_{\textbf{k}} \equiv \delta \sigma_{\textbf{k}}/a$~, one has
\begin{eqnarray}
\label{deltaphik}
\left|\delta \phi_{\textbf{k}}\right| &=& \left|\frac{1}{a}\cdot \frac{e^{-i\sqrt{k^2+\varepsilon^{*2}_{\gamma}}~\tau}}{\sqrt{2}\left(k^2+\varepsilon^{*2}_{\gamma}\right)^{1/4}}
\left(1+ \frac{i}{\sqrt{k^2+\varepsilon^{*2}_{\gamma}}~\tau}\right)\right| \nonumber \\
&=& \frac{1}{\sqrt{2}a \left(k^2+\varepsilon^{*2}_{\gamma}\right)^{1/4}} \sqrt{1+ \frac{1}{\left(k^2+\varepsilon^{*2}_{\gamma}\right)~\tau^2}}\ .
\end{eqnarray}

\begin{figure}
\resizebox{0.43\textwidth}{!}{
\includegraphics{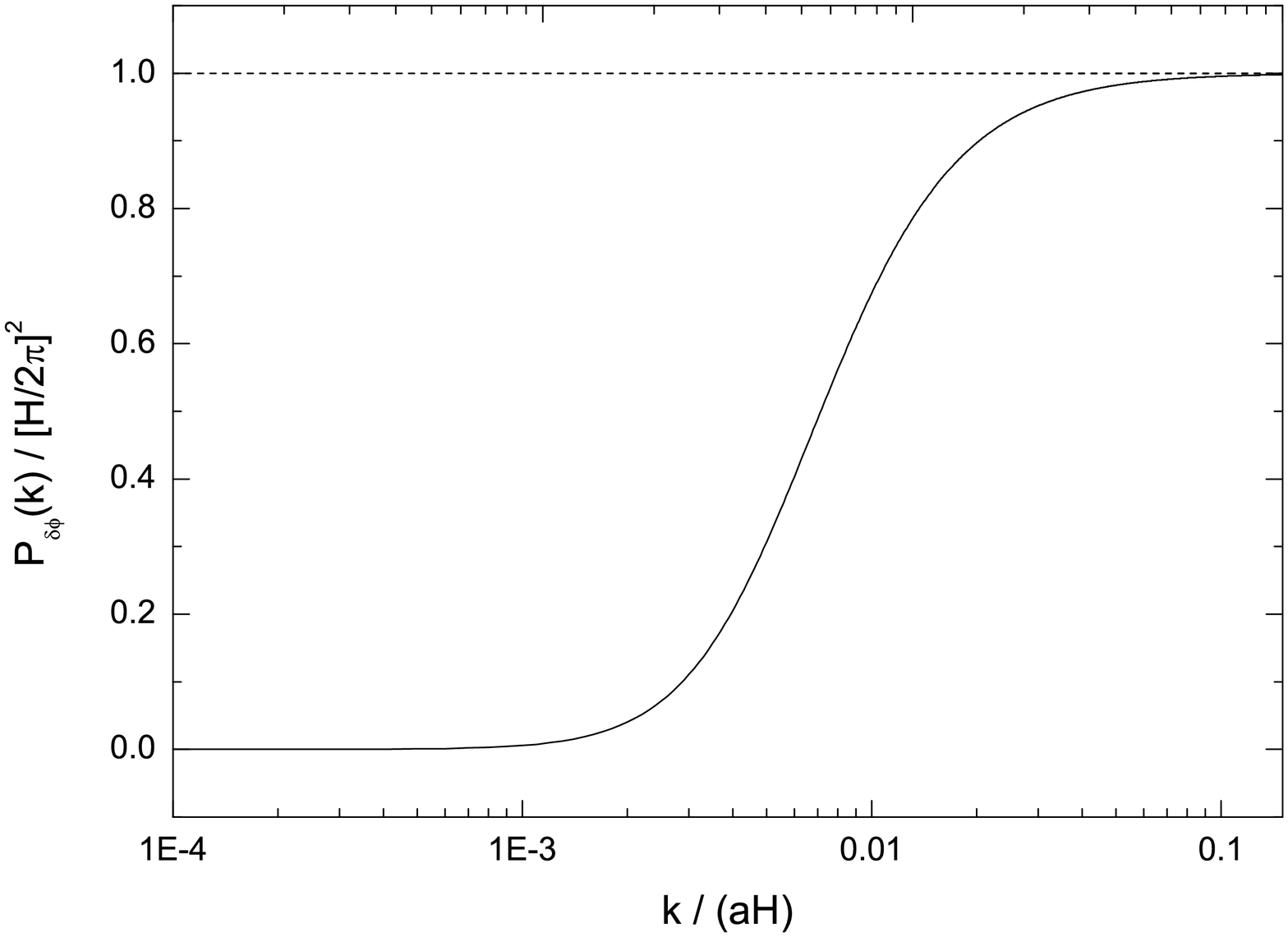}}
\caption{The non-dimensional primordial power spectrum ${\cal P}_{\delta \phi}(k)$. It experiences a cutoff at large scales for certain values of $\varepsilon^{*}_{\gamma}$. It is also shown in the bottom panel of Fig.\ref{fig2}. }
\label{fig1}
\end{figure}

The power spectrum of $\delta \phi_{\textbf{k}}$, which is denoted by ${\cal P}_{\delta \phi}(k)$, is defined as \cite{A2002-}
\begin{equation}
\langle 0|\delta \phi^{*}_{{\bf k}_1}\delta \phi_{{\bf k}_2}|0\rangle \equiv \delta^{(3)}\left({\bf k}_1-{\bf k}_2\right)\,\frac{2\pi^2}{k^3}\,{\cal P}_{\delta \phi}(k)\ .
\label{powerdef0}
\end{equation}
For $\textbf{k}_1 =\textbf{k}_2 = \textbf{k}$, one has
\begin{equation}
\label{powerdef}
{\cal P}_{\delta \phi}(k) \equiv \frac{k^3}{2\pi^2}\,|\delta \phi_{\textbf{k}}|^2\ .
\end{equation}
Given the results (\ref{deltaphik}) and  $a=-1/H\tau$ \cite{A2002-}, one obtains the primordial power spectrum of the perturbation $\delta\phi({\bf x},t)$,
\begin{eqnarray}
\label{powerdeltaphi}
{\cal P}_{\delta \phi}(k) &=& \frac{k^3}{2\pi^2}\cdot\frac{1}{2a^2 \sqrt{k^2+\varepsilon^{*2}_{\gamma}}} \left(1+ \frac{1}{\left(k^2+\varepsilon^{*2}_{\gamma}\right)~\tau^2}\right)\nonumber \\
&=& \frac{H^2}{4\pi^2}\cdot \frac{k^3}{\left({k^2+\varepsilon^{*2}_{\gamma}}\right)^{3/2}}
+ \frac{H^2}{4\pi^2}\cdot \frac{k}{\sqrt{k^2+\varepsilon^{*2}_{\gamma}}}\cdot k^2\tau^2\ .
\end{eqnarray}
According to the super-horizon criterion, the wavelength $\lambda$ of the primordial perturbation is beyond the horizon if
\begin{equation}
\label{ktau}
-k\tau = \frac{k}{aH} \ll 1\ .
\end{equation}
Therefore, given the super-horizon condition (\ref{ktau}), the second term at the right hand side of the expression (\ref{powerdeltaphi}) becomes sufficiently small when compared to the first term and can be neglected. The power spectrum is then given approximately as
\begin{equation}
\label{powerdeltaphi1}
{\cal P}_{\delta \phi}(k) \simeq \frac{H^2}{4\pi^2}\cdot \frac{k^3}{\left({k^2+\varepsilon^{*2}_{\gamma}}\right)^{3/2}}\ .
\end{equation}
For large $k$ (i.e. high multipole index $\ell$), one recovers the usual scale-invariant primordial power spectrum of the perturbation
\begin{equation}
{\cal P}_{\delta \phi}(k) =  \frac{H^2}{4\pi^2}\cdot \frac{k^3}{\left[k^2 \left(1+\frac{\varepsilon^{*2}_{\gamma}}{k^2}\right)\right]^{3/2}}
\simeq \frac{H^2}{4\pi^2}\cdot \frac{k^2}{k^2 + \frac{3}{2}\varepsilon^{*2}_{\gamma}}~\sim~k^0\, .
\end{equation}
Plot of the primordial spectrum (\ref{powerdeltaphi1}) is shown in Fig.\ref{fig1}.

\section{The Resulting CMB Spectrum}
\label{sec:3}
The temperature fluctuations of the CMB are due to the quantum fluctuations of a weakly self-coupled scalar matter field $\phi$ , which later results in an exponential inflation of the early universe \cite{Star}-\cite{Mukh2}. The anisotropic and inhomogeneous fluctuations $\delta\phi(\textbf{x},t)$ of this scalar field result in the perturbations of the comoving curvature $\mathcal{R}$. During inflation, the wavelengths $\lambda$ of these perturbations are then stretched exponentially out of the horizon $\mathcal{H}$. As time goes by, the horizon of the universe grows. At some time after inflation, these once ``frozen'' perturbations finally reenter the horizon and causal region. For the rest of the time, they evolve according to the Poisson equation and the collisionless Boltzmann equation, giving rise to the extant matter and temperature fluctuations. The primordial perturbation of photons at the surface of last scattering is the one that responsible for the anisotropic power spectrum of the CMB radiation we observe today.

The power spectrum of the comoving curvature perturbation $\mathcal{R}$ is usually denoted by $\mathcal{P}_{\mathcal{R}}(k)$, which is given as \cite{A2002-}
\begin{equation}
\label{powerdef2}
\mathcal{P}_{\mathcal{R}}(k)  = \frac{H^2}{\dot{\phi}^2}~{\cal P}_{\delta \phi}(k) \equiv A^2_s \left(\frac{k}{aH} \right)^{n_s-1}\ ,
\end{equation}
where $A_s$ is the normalized amplitude and $n_s$ is the scalar spectral index. In the slow roll inflation model, $n_s-1=2\eta-6\epsilon$.
The anisotropy spectrum of the CMB, indicated by the coefficient $C_{\ell}$ , is obtained by a line-of-sight integration over the spectrum $\mathcal{P}_{\mathcal{R}}(k)$ with $\Delta_{\ell}(k,\tau)$, which is the solution of the collisionsless Boltzmann equation for the CMB photons, to wit \cite{C1995-}\cite{Seljak}
\begin{equation}
C_{\ell}=4\pi \int d^3k~\mathcal{P}_{\mathcal{R}}(k)|\Delta_{\ell}(k,\tau)|^2\ .
\label{cl}
\end{equation}
The solution $\Delta_{\ell}(k,\tau)$ is obtained by numerically solving the coupled Boltzmann equations of the CMB photons under the adiabatic initial conditions \cite{Bucher2000-}.

\begin{figure}
\resizebox{0.44\textwidth}{!}{
\includegraphics{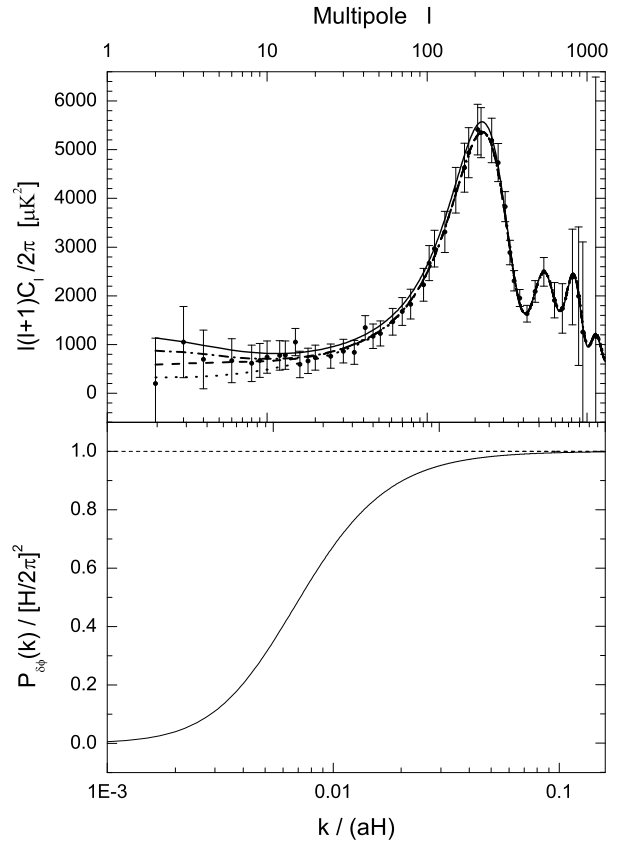}}
\caption{The CMB angular power spectrum $C_{\ell}$ vs.$~{\ell}$ with the primordial power spectrum ${\cal P}_{\delta \phi}(k)$ vs.$~k/(aH)$ . Black dots with error bars represent (part of) the TT data of the WMAP 7-year results. The black solid curve in the top panel is the theoretical prediction of the standard $\Lambda$-CDM model with the usual scale-invariant primordial power spectrum ${\cal P}_{\delta \phi}(k) = H^2/4\pi^2$ (i.e. the spectral index $n_\textmd{s}= 1$), and with the cosmological parameters fixed at $h=0.73$, $\Omega_\textmd{b} h^2=0.0226$, $\Omega_{\textmd{cdm}} h^2= 0.112$, $\Omega_\Lambda=0.728$, $\tau =0.087$. The dashed and dotted curves are obtained with the same model, but with the infrared-cutoff primordial power spectrum (\ref{powerdeltaphi1}). From up to down each of them sequently corresponds to the theoretical result with $\varepsilon^{*}_{\gamma}= 7\times 10^{-5}$(dash-dotted), $2.5\times 10^{-4}$(dashed), and $7.6\times 10^{-4}$(dotted) $\textmd{Mpc}^{-1}$. For $\varepsilon^{*}_{\gamma}=2.5\times 10^{-4}$ $\textmd{Mpc}^{-1}$, the cosmic age is $13.79$ Gyr and the value of $\sigma_8$ is $0.809$.The asymptotic behavior of the primordial spectrum ${\cal P}_{\delta \phi}(k)$ at large scales is shown in the bottom panel.}
\label{fig2}
\end{figure}

\begin{figure}
\resizebox{0.44\textwidth}{!}{\includegraphics{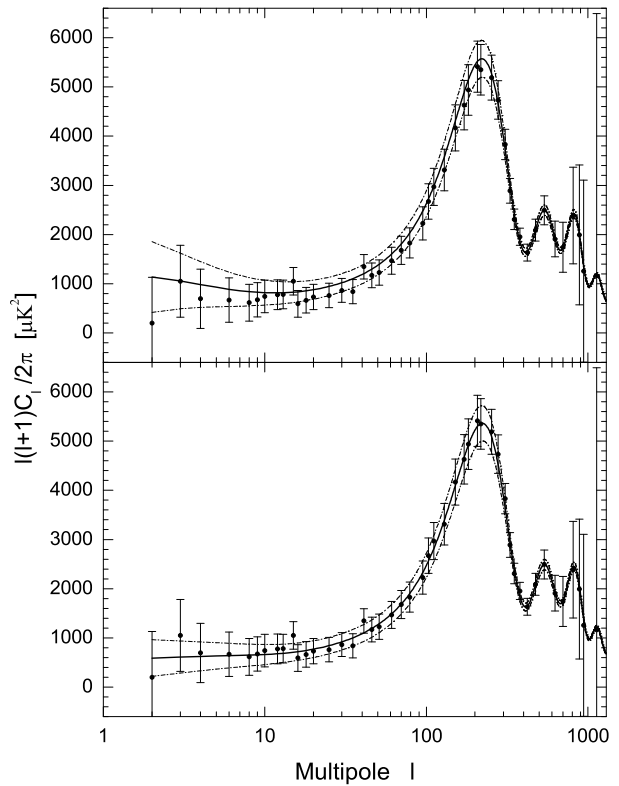}}
\caption{The CMB angular power spectrum $C_{\ell}$ vs. ${\ell}$ . Black dots with error bars in the two panels represent (part of) the TT data of the WMAP-7 results. The contributions of the cosmic variance to the errors in the two models, the standard one and ours, are shown by the dash-dotted upper and lower contours in both panels. In the top panel, the black solid curve is the theoretical prediction of the standard $\Lambda$-CDM model with the usual scale-invariant primordial power spectrum ${\cal P}_{\delta \phi}(k) = H^2/4\pi^2$ (i.e. the spectral index $n_s= 1$), and with the cosmological parameters fixed at $h=0.73$, $\Omega_\textmd{b} h^2=0.0226$, $\Omega_{\textmd{cdm}} h^2= 0.112$, $\Omega_\Lambda=0.728$, $\tau =0.087$. In the bottom panel, the curves are obtained with the same cosmological model, but with the modified power spectrum (\ref{powerdeltaphi1}) when $\varepsilon^{*}_{\gamma}=2.5\times 10^{-4}$ $\textmd{Mpc}^{-1}$. The cosmic age is $13.79$ Gyr and the value of $\sigma_8$ is $0.809$.}
\label{fig3}
\end{figure}

We use a modified version of the publicly available Code for Anisotropies in the Microwave Background(CAMB) \cite{CAMB} to compute the CMB temperature-temperature(TT) spectra with various $\varepsilon^{*}_{\gamma}$ in the modified primordial spectrum (\ref{powerdeltaphi1}). The numerical results are shown in Fig.\ref{fig2} and Fig.\ref{fig3}. The black solid curve in Fig.\ref{fig2} indicates the standard scale-invariant primordial spectrum model with the same cosmological parameters as those obtained from the best-fit to the WMAP 7-year data. The dashed and dotted curves represent the theoretical predictions of our model. The cosmic variance contributions to the errors in both models of inflation (the standard one and ours) are shown respectively by the dash-dotted upper and lower contours in the two panels of Fig.\ref{fig3}. For a simple numerical analysis, we arbitrarily choose the value of $\varepsilon^{*}_{\gamma}$ to be $7\times 10^{-5}$, $2.5\times 10^{-4}$, and $7.6\times 10^{-4}$ $\textmd{Mpc}^{-1}$. The corresponding $\chi^2$ are respectively $1530.83063$, $1534.18154$, and $1560.87900$, while the standard inflation model gives $\chi^2=1380.16810$ (in both cases for $1200$ data points).
For $\varepsilon^{*}_{\gamma}= 2.5\times 10^{-4}$ $\textmd{Mpc}^{-1}$, the age of the universe is $13.79$ Gyr and the value of $\sigma_8$ is $0.809$. Both of them are consistent with the values presented in \cite{pdg2009}.

From Fig.\ref{fig2}, we see that the lower CMB quadrupole is related to the value of $\varepsilon^{*}_{\gamma}$. It implies that in some sense the lower CMB quadrupole encodes the information of the geometric properties of the de Sitter space. A zero $\varepsilon^{*}_{\gamma}$ will land us back at the standard scale-invariant spectrum model.
In order to obtain a more definite result, more data such as COBE \cite{COBE}, BOOMERANG \cite{Netterfield01}, MAXIMA \cite{Hanany00}, DASI \cite{Halverson01}, VSA \cite{VSA3} and CBI \cite{Pearson02}\cite{CBIdata} are needed for further numerical studies. Careful Monte Carlo simulations or a full Bayesian analysis is also necessary for drawing a more confirmatively quantitative conclusion \cite{Lewis2002-Monte}. But these are out of our research area. A simple numerical-recipe-level analysis may be enough for taking a look at possible qualitative features of the solution (\ref{powerdeltaphi1}).

\section{Conclusions and Discussions}
\label{sec:4}
In this paper, we studied the possible effects of cosmological-constant type dark energy on the standard inflationary paradigm of modern cosmology. We presented a unified scenario of the the $\Lambda$-CDM and the inflation model. This is accommodated by the new dispersion relation (\ref{dispersion1}). It stems from the kinematics of free particles in a four dimensional de Sitter space. We got a modified inflation model of a minimal coupled scalar field. The ultraviolet behavior of the primordial spectrum in our model differs little from the usual scale-invariant one, which ensures an agreement with the WMAP 7-year observations for high-$\ell$ components.
For certain values of the model parameter $\varepsilon^{*}_{\gamma}$, the model is able to generate a power spectrum with lower energy on large scales.
And from the relation (\ref{dispersion1}) and $R=1/\sqrt{K}$, we approximately have $R\sim 10^{3}$ $\textmd{Mpc}$ for $\varepsilon^{*}_{\gamma}\sim 10^{-4}$ $\textmd{Mpc}^{-1}$. If one agrees that our universe is asymptotic to the Robertson-Walker-like de Sitter space of $R\simeq (3/\Lambda )^{1/2}$ \cite{Guo2008-}, from $\Omega_{\Lambda}\equiv \Lambda / 3H_0^2$ one finally obtains $\Omega_{\Lambda} \sim 10^{-1}$. This is in reasonable agreement with the current acknowledged value of $\Omega_{\Lambda}\simeq 0.72$ \cite{pdg2009}.

We note, however, that as pointed out by C. Bennett \textit{et al.} \cite{Bennett2011}, the mean value of the CMB quadrupole component $C_2$ predicted by the best-fit $\Lambda$-CDM model lies within the $95\%$ confidence region allowed by the data. If this is true, that means the measured value of the quadrupole is not anomalously low. So one has to take this rough model of inflation with a few grains of salt.
In this paper, we just offered a theoretical possibility --- the inflationary paradigm of a single scalar field with $\Lambda$-type dark energy is still able to yield an observation-compatible scale-invariant primordial spectrum, while having the potential to generate a spectrum with low-$\ell$ multipoles. The result we obtained is undoubtedly sketchy. As stated at the end of Section IV, a more careful and comprehensive numerical analysis like that done by the WMAP team should be carried out before drawing a confirmative quantitative conclusion. But that requires professional data analysis techniques which is out of our research field.


Last, a close relation between de Sitter space and Doubly Special Relativity(DSR) should be noticed. DSR, first proposed by Amelino-Camelia around the start of this millenium \cite{amelino2000-}, is one of the possible explanations of the GZK feature in the energy spectrum of the ultra-high energy cosmic rays without invoking Lorentz symmetry violation. It is based on two fundamental assumptions:
\begin{itemize}
\item  \textit{The principle of relativity still holds, i.e., all the inertial observers are equivalent};
\item  \textit{There exists two observer-independent scales: one is of dimension of velocity, which is identified with the speed of light $c$; the other is of dimension of length $\kappa$ (or mass $\kappa^{-1}$), which is identified with the Planck length (or mass)}.
\end{itemize}

The energy-momentum space of DSR is found to be a four dimensional maximally symmetric one. In differential geometry, such a manifold must be locally diffeomorphic to one of the three kind of spaces of constant curvature: \emph{de Sitter}, \emph{Minkowski}, and \emph{Anti-de Sitter}, of which the sign of curvature $K$ is respectively $+$, $0$, and $-$. It was first pointed out by J. Kowalski-Glikman \cite{J2002-}\cite{gac} and later reaffirmed by H.-Y. Guo \cite{Guo2007-} that DSR can be taken as a theory with its energy-momentum space being a four dimensional de Sitter space. Different formulations of DSR can be identified with taking different coordinate systems on this space. In our paper, the curvature radius $R$ of the de Sitter space plays the role of the length scale $\kappa$ in DSR and our research is formulated in the Beltrami coordinates.

However, more research, in the theoretical as well as the phenomenological aspects, needs to be done before claiming that the particle kinematics in DSR can be identified with that in a de Sitter space. Relevant researches are currently undertaken.

%

\begin{thebibliography}{}
%
%

\bibitem{wmap7}
N.~Jarosik {\it et al.}, ``Seven-Year Wilkinson Microwave Anisotropy Probe (WMAP) Observations: Sky Maps, Systematic Errors, And Basic Results,''  Astrophys. J. Suppl. {\bf 192}, 14 (2011). (The WMAP 7-year data is publicly available on the website \texttt{http://lambda.gsfc.nasa.gov/product/map/current/m\_products.cfm}~)


\bibitem{0067-0049-192-2-18}
E.~Komatsu {\it et al.}, ``Seven-Year Wilkinson Microwave Anisotropy Probe (WMAP) Observations:
  Cosmological Interpretation,'' Astrophys. J. Suppl. {\bf 192}, 18 (2011).


\bibitem{Star} A.~A.~Starobinsky,
``Spectrum Of Relict Gravitational Radiation And The Early State
Of The Universe,'' JETP Lett.\  {\bf 30}, 682 (1979) [Pisma Zh.\
Eksp.\ Teor.\ Fiz.\  {\bf 30}, 719 (1979)]; A.~A.~Starobinsky, ``A
New Type Of Isotropic Cosmological Models Without Singularity,''
Phys.\ Lett.\ B {\bf 91}, 99 (1980).


\bibitem{Mukh} V.~F.~Mukhanov and G.~V.~Chibisov,
``Quantum Fluctuation And `Nonsingular' Universe,'' JETP Lett.\
{\bf 33}, 532 (1981) [Pisma Zh.\ Eksp.\ Teor.\ Fiz.\  {\bf 33},
549 (1981)].


\bibitem{Guth} A.~H.~Guth,
``The Inflationary Universe: A Possible Solution To The Horizon
And Flatness Problems,'' Phys.\ Rev.\ D {\bf 23}, 347 (1981).


\bibitem{New}
A.~D.~Linde, ``A New Inflationary Universe Scenario: A Possible
Solution Of The Horizon, Flatness, Homogeneity, Isotropy And
Primordial Monopole Problems,'' Phys.\ Lett.\ B {\bf 108}, 389
(1982); A.~D.~Linde,
``Coleman-Weinberg Theory And A New Inflationary Universe Scenario,''
  Phys.\ Lett.\ B {\bf 114}, 431 (1982);
A.~D.~Linde,
``Temperature Dependence Of Coupling Constants And The Phase Transition In
The Coleman-Weinberg Theory,''
  Phys.\ Lett.\ B {\bf 116}, 340 (1982);
A.~D.~Linde,
``Scalar Field Fluctuations In Expanding Universe And The New Inflationary
Universe Scenario,''
  Phys.\ Lett.\ B {\bf 116}, 335 (1982).

\bibitem{New2} A.~Albrecht and P.~J.~Steinhardt, ``Cosmology For Grand
Unified Theories With Radiatively Induced Symmetry Breaking,''
Phys.\ Rev.\ Lett.\  {\bf 48}, 1220 (1982).


\bibitem{Hawk} S.~W.~Hawking,
``The Development Of  Irregularities  In  A Single Bubble
Inflationary Universe,'' Phys.\ Lett.\ B {\bf 115}, 295 (1982);
A.~A.~Starobinsky, ``Dynamics Of Phase Transition In The New
Inflationary Universe Scenario And Generation Of Perturbations,''
Phys.\ Lett.\ B {\bf 117}, 175 (1982); A.~H.~Guth and S.~Y.~Pi,
``Fluctuations In The New Inflationary Universe,'' Phys.\ Rev.\
Lett.\  {\bf 49}, 1110 (1982); J.~M.~Bardeen, P.~J.~Steinhardt and
M.~S.~Turner, ``Spontaneous Creation Of Almost Scale - Free
Density Perturbations In An Inflationary Universe,'' Phys.\ Rev.\
D {\bf 28}, 679 (1983).

\bibitem{Mukh2} V.~F.~Mukhanov,
``Gravitational Instability Of The Universe Filled With A Scalar
Field,'' JETP Lett.\  {\bf 41}, 493 (1985) [Pisma Zh.\ Eksp.\
Teor.\ Fiz.\  {\bf 41}, 402 (1985)]; V.~F.~Mukhanov, H.~A.~Feldman
and R.~H.~Brandenberger, ``Theory Of Cosmological Perturbations,''
Phys.\ Rept.\  {\bf 215}, 203 (1992); V. F. Mukhanov, {\it Physical Foundations of Cosmology}, Cambridge University Press, 2008.



\bibitem{H2010-}
H.~Kurki-Suonio, ``Physics Of The Cosmic Microwave Background And The Planck Mission,'' Proceedings of the 2010 CERN Summer School, Raseborg (Finland), submitted for publication in a CERN Yellow Report (2010) [arXiv:1012.5204v1].

\bibitem{Larson2011-}
D.~Larson {\it et al.}, ``Seven-year Wilkinson Microwave Anisotropy Probe (WMAP) Observations: Power Spectra And WMAP-Derived Parameters,'' Astrophys. J. Suppl. {\bf 192}, 16 (2011).


\bibitem{J2003-small}
J.~M.~Cline, P.~Crotty and J.~Lesgourgues, ``Does The Small CMB Quadrupole Moment Suggest New Physics?,'' J. Cosmol. Astropart. P. {\bf 9}, (2003).

\bibitem{Efst}
G.~Efstathiou, ``Is The Low CMB Quadrupole A Signature Of Spatial Curvature?,'' Mon.~Not.~Roy.~Astron.~Soc. {\bf 343}, L95 (2003) [arXiv:astro-ph/0303127].

\bibitem{Kurki-Suonio2010}
H. Kurki-Suonio, ``Physics of the Cosmic Microwave Background and the Planck Mission,'' Proceedings of the 2010 CERN Summer School, Raseborg (Finland), submitted for publication in a CERN Yellow Report [arXiv:1012.5204v1].


\bibitem{Jing}
Y.~P.~Jing and L.~Z.~Fang,
``An Infrared Cutoff Revealed By The Two Years Of COBE - DMR Observations Of Cosmic Temperature Fluctuations,''
Phys.\ Rev.\ Lett.\  {\bf 73}, 1882 (1994) [arXiv:astro-ph/9409072].

\bibitem{Scott2010}
D.~Scott and G.~F.~Smoot, ``Cosmic Microwave Background Mini-review,'' appearing in the 2010 Review of Particle Physics, available on the PDG website at [\texttt{http://pdg.lbl.gov/2011/astrophysics-cosmology/astro-cosmo.html or http://pdg.lbl.gov/2011/reviews/rpp2011-rev-cosmic-microwave-background.pdf}]


\bibitem{Li2010-}
H.~Liu, S.-L.~Xiong, and T.-P.~Li, ``The Origin Of The WMAP Quadrupole,'' (2010) [arXiv:1003.1073v2].

\bibitem{Li2011-}
H.~Liu and T.-P.~Li, ``Observational Scan Induced Artificial CMB Anisotropy,''  Astrophys. J. (2011), to be published [arXiv:1003.1073v2].



\bibitem{Kowaski2003-}
M.~Kawasaki and F.~Takahashi, ``Inflation Model With Lower Multipoles Of The CMB Suppressed,'' Phys. Lett. B {\bf 570}, 151 (2003).

\bibitem{Feng2003-}
B.~Feng and X.-M.~Zhang, ``Double Inflation And The Low CMB Quadrupole,'' Phys. Lett. B {\bf 570}, 145 (2003).

\bibitem{C2003-Suppressing}
C.~R.~Contaldi \textit{et~al.}, ``Suppressing The Lower Multipoles In The CMB Anisotropies,'' J. Cosmol. Astropart. P. {\bf 9}, (2003).

\bibitem{Y2005-Possible}
Y.-S.~Piao, ``Possible Explanation To A Low CMB Quadrupole,'' Phys.\ Rev.\ D {\bf 71}, 087301 (2005).

\bibitem{Kofman:1989ed}
L.~Kofman, G.~R.~Blumenthal, H.~Hodges and J.~R.~Primack,
``Generation Of Nonflat And Nongaussian Perturbations From Inflation,''
ASP Conf.\ Ser.\  {\bf 15}, 339 (1991).

\bibitem{Lidsey:1995np}
J.~E.~Lidsey, A.~R.~Liddle, E.~W.~Kolb, E.~J.~Copeland, T.~Barreiro and M.~Abney,
``Reconstructing The Inflaton Potential--An overview,''
Rev.\ Mod.\ Phys.\  {\bf 69}, 373 (1997)
[arXiv:astro-ph/9508078].

\bibitem{Hodges:bf}
H.~M.~Hodges and G.~R.~Blumenthal,
``Arbitrariness Of Inflationary Fluctuation Spectra,''
Phys.\ Rev.\ D {\bf 42}, 3329 (1990).

\bibitem{Arkani-Hamed:2002fu}
N.~Arkani-Hamed, S.~Dimopoulos, G.~Dvali and G.~Gabadadze,
``Non-Local Modification of Gravity And The Cosmological Constant Problem,''
[arXiv:hep-th/0209227];
N.~Arkani-Hamed, S.~Dimopoulos, G.~Dvali, G.~Gabadadze, and A.D. Linde,
``Self-Terminating Inflation,''
in preparation.

\bibitem{Linde:1994gy}
A.~D.~Linde, D.~A.~Linde and A.~Mezhlumian,
``Do We Live In The Center Of The World?,''
Phys.\ Lett.\ B {\bf 345}, 203 (1995)
[arXiv:hep-th/9411111].
A.~D.~Linde, D.~A.~Linde and A.~Mezhlumian,
``Nonperturbative Amplifications Of Inhomogeneities In A Self-Reproducing Universe,''
Phys.\ Rev.\ D {\bf 54}, 2504 (1996)
[arXiv:gr-qc/9601005].

\bibitem{Starobinsky:ts}
A.~A.~Starobinsky,
``Spectrum Of Adiabatic Perturbations In The Universe When There Are Singularities In The Inflation Potential,''
JETP Lett.\  {\bf 55}, 489 (1992)
[Pisma Zh.\ Eksp.\ Teor.\ Fiz.\  {\bf 55}, 477 (1992)].

\bibitem{Hodges:1989dw}
H.~M.~Hodges, G.~R.~Blumenthal, L.~A.~Kofman and J.~R.~Primack,
``Nonstandard Primordial Fluctuations From A Polynomial Inflaton Potential,''
Nucl.\ Phys.\ B {\bf 335}, 197 (1990).

\bibitem{Rodrigues2008-}
D.~C.~Rodrigues, ``Anisotropic Cosmological Constant And The CMB Quadrupole Anomaly,'' Phys.\ Rev.\ D {\bf 77}, 023534 (2008).


\bibitem{Copi2010}
C.~J.~Copi, ``Large-angle Anomalies in the CMB,'' Advances in Astronomy {\bf 2010}, 847541 (2010).



\bibitem{Bennett2011}
C.~L. Bennett {\it et al.}, ``Seven-Year Wilkinson Microwave Anisotropy Probe (Wmap) Observations: Are There Cosmic Microwave Background Anomalies?'' Astrophys. J. Suppl. {\bf 192}, 17 (2011).



\bibitem{Ho}
R.~H.~Brandenberger and P.~M.~Ho,
``Noncommutative Spacetime, Stringy Spacetime Uncertainty Principle, and Density Fluctuations,''
Phys.\ Rev. D {\bf 66}, 023517 (2002)
[arXiv:hep-th/0203119].


\bibitem{Brand2003-}
R.~H. Brandenberger, ``Trans-Planckian Physics And Inflationary Cosmology,'' Proceedings of the 2002 International Symposium on Cosmology and Particle Astrophysics, 101 (2003) [arXiv:hep-th/0210186v2].

\bibitem{J2001-Trans-Planckian}
J.~Martin and R.~H. Brandenberger, ``Trans-Planckian Problem Of Inflationary Cosmology,''
Phys.\ Rev.\ D {\bf 63}, 123501 (2001).




\bibitem{Peebles2003-}
P.~J.~E. Peebles and B.~Ratra, ``The Cosmological Constant And Dark Energy,'' Rev. Mod. Phys. {\bf 75}, 559 (2003) [arXiv:astro-ph/0207347v2].


\bibitem{S2004-Cosmic}
Z.~Chang, S.-X.~Chen and C.-B. Guan, ``Cosmic Ray Threshold Anomaly And Kinematics In The dS Spacetime,'' (2004) [arXiv:astro-ph/0402351v1]; Z.~Chang, S.-X.~Chen, C.-B. Guan and C.-G. Huang, ``Cosmic Ray Threshold In An Asymptotically DS Spacetime,'' Phys. Rev. D {\bf 71}, 103007 (2005) [arXiv:astro-ph/0505612v1].





\bibitem{G2003-Beltrami}
H.-Y. Guo, C.-G. Huang, Z.~Xu and B.~Zhou, ``On Beltrami Model Of De Sitter Spacetime,'', Mod.~Phys.~Lett. A {\bf 19}, 1701 (2004).




\bibitem{A2002-}A.~Riotto, ``Inflation And The Theory Of Cosmological Perturbations,'' Lectures given at the: Summer School on Astroparticle Physics and Cosmology (2002) [arXiv:hep-ph/0210162v1].

\bibitem{Dodelson2003-}
S.~Dodelson, \textit{Modern Cosmology}, Elsevier(Singapore) Pte Ltd., 2003.



\bibitem{C1995-}
C.-P.~Ma and E.~Bertschinger, ''Cosmological Perturbation-Theory In The Synchronous And Conformal Newtonian Gauges,'' Astrophys. J. {\bf 455}, 7 (1995) [arXiv:astro-ph/9506072v1].

\bibitem{Seljak}
U.~Seljak and M.~Zaldarriaga, ``A Line Of Sight Approach To Cosmic Microwave Background Anisotropies,'' Astrophys. J. {\bf 469}, 437 (1996) [arXiv:astro-ph/9603033].

\bibitem{Bucher2000-}
M.~Bucher, K.~Moodley and N.~Turok, ``The General Primordial Cosmic Perturbation,'' Phys. Rev. D {\bf 62}, 083508 (2000).

\bibitem{CAMB}
A.~Lewis and A.~Challinor, Code for Anisotropies in the Microwave Background(CAMB) (2011) [\texttt{http://camb.info/}] (This code is based on CMBFAST by U.~Seljak and M.~Zaldarriaga (1996) [\texttt{http://lambda.gsfc.nasa.gov/toolbox/tb\_ cmbfast\_ ov.cfm}]).

\bibitem{pdg2009}
O.~Lahav and A.~Liddle, ``The Cosmological Parameters,'' appearing in the 2010 Review of Particle Physics, available on the PDG website at [\texttt{http://pdg.lbl.gov/2011/astrophysics-cosmology/astro-cosmo.html} or \texttt{http://pdg.lbl.gov/2011/reviews/rpp2011-rev-cosmic-microwave-background.pdf}].


\bibitem{COBE}
C. L. Bennett \textit{et al.}, ``Four-Year COBE* DMR Cosmic Microwave Background Observations: Maps And Basic Results,'' Astrophys. J. Lett. {\bf 464}, (1996).

\bibitem{Netterfield01}
C. B. Netterfield \textit{et al.}, ``A Measurement By BOOMERANG Of Multiple Peaks In The Angular Power Spectrum Of The Cosmic Microwave Background,'' Astrophys. J. {\bf 571}, 604 (2002) [arXiv:astro-ph/0104460].

\bibitem{Hanany00}
S. Hanany \textit{et al.}, ``MAXIMA-1: A Measurement Of The Cosmic Microwave Background Anisotropy On Angular Scales of 10 Arcminutes To 5 Degrees
,'' Astrophys. J. {\bf 545}, L5 (2000) [arXiv:astro-ph/0005123].

\bibitem{Halverson01}
N. W. Halverson \textit{et al.}, ``DASI First Results: A Measurement Of The Cosmic Microwave Background Angular Power Spectrum
,'' Astrophys. J. {\bf 568}, 38 (2002) [arXiv:astro-ph/0104489].

\bibitem{VSA3}
P. F. Scott \textit{et al.}, ``First Results From The Very Small Array---III. The CMB power spectrum,'' Mon. Not. Roy. Astron. Soc. {\bf 341}, 1076 (2003) [arXiv:astro-ph/0205380].

\bibitem{Pearson02}
T. J. Pearson \textit{et al.}, ``The Anisotropy of the Microwave Background To $\ell$ = 3500: Mosaic Observations With The Cosmic Background Imager
,'' Astrophys. J. {\bf 591}, 556 (2003) [arXiv:astro-ph/0205388].

\bibitem{CBIdata}
CBI Supplementary Data, 10 July (2002) [\texttt{http://www.astro.caltech.edu/$\sim$tjp/CBI/data/}].

\bibitem{Lewis2002-Monte}
A.~Lewis and S.~Bridle, ``Cosmological Parameters From CMB And Other Data: A Monte-Carlo Approach,'' Phys. Rev. D {\bf 66}, 103511 (2002) [arXiv:astro-ph/0205436].





\bibitem{Guo2008-}
H.~Y.~Guo, Science in China A, {\bf 51}, 568 (2008).



\bibitem{amelino2000-}
G.~Amelino-Camelia, ``Relativity In Space-times With Short-Distance Structure Governed By An Observer-Independent (Planckian) Length Scale,'' Int.J.Mod.Phys. D {\bf 11}, 35 (2002) [arXiv:gr-qc/0012051]; G.~Amelino-Camelia, ``Testable Scenario For Relativity With Minimum-Length,'' Phys.~Lett. B {\bf 510}, 255, (2001) [arXiv:hep-th/0012238]; G.~Amelino-Camelia, ``Doubly Special Relativity,'' Nature {\bf 418}, 34 (2002) [arXiv:gr-qc/0207049].


\bibitem{J2002-}
J.~Kowalski-Glikman, ``De Sitter Space As An Arena For Doubly Special Relativity,'' Phys. Lett. B {\bf 547}, 291 (2002).

\bibitem{gac}
J.~Kowalski-Glikman, ``Introduction To Doubly Special Relativity,'' Lect. Notes Phys. {\bf 669}, 131 (2005).


\bibitem{Guo2007-}
H.~Y.~Guo \textit{et al.}, ``Snyder's Model¡ª- De Sitter Special Relativity Duality And De Sitter Gravity'', Class. Quantum Grav. {\bf 24}, 4009 (2007).



\end{thebibliography}
%

\end{document}